\documentclass[aps,prb,twocolumn,floatfix,showpacs]{revtex4}
\usepackage{graphicx}
\usepackage{amsmath,bm,amssymb,amsfonts}

\begin{document}

\title{Magnetic and transport signatures of Rashba spin-orbit coupling
on the ferromagnetic Kondo lattice model in two dimensions.}

\author{Giovany A. Meza and Jos\'e A. Riera}
\affiliation{Instituto de F\'{\i}sica Rosario y Departamento de
F\'{\i}sica,\\
Universidad Nacional de Rosario, Rosario, Argentina
}

\date{\today}

\begin{abstract}
Motivated by emergent phenomena at oxide surfaces and interfaces,
particularly those involving transition metal oxides with perovskite
crystal structure such as LaTiO$_3$/SrTiO$_3$, we examine the 
ferromagnetic Kondo lattice model (FKLM) in the presence of a Rashba
spin-orbit coupling (RSOC).
Using numerical techniques, under the assumption that the electrons on
localized orbitals may be treated as classical continuum spins, we
compute various charge, spin and transport properties on square clusters
at zero temperature. We find that the main effect of the RSOC is the 
destruction of the ferromagnetic state present in the FKLM at low 
electron fillings, with the consequent suppression of conductivity. In
addition, near half-filling the RSOC leads to a departure of the
antiferromagnetic state of the FKLM with a consequent reduction to
the intrinsic tendency to electronic phase separation.
The interplay between phase separation on one side, and magnetic and
transport properties on the other, is carefully analyzed as a function
of the RSOC/hopping ratio.
\end{abstract}


\maketitle

\section{Introduction}

Transition metal oxides (TMOs) are typical strongly correlated systems
because of the available d-orbitals, and are characterized by a complex
interrelation between charge, spin and orbital degrees of 
freedom, leading in turn to remarkable properties such as high-T$_c$ 
superconductivity in cuprates and colossal magnetoresistance in 
manganites.
The intensive study of these strongly correlated systems has identified
the ultimate origin of those phenomena as caused by the collective
behaviour of electrons, which could only be captured by appropriate
many-body approaches.

Additional emergent phenomena at the interface between strongly
correlated materials, and particularly TMOs, or at the surface of such
materials, have been revealed even more recently by a number of
theoretical and experimental studies.\cite{hwang}
In essence, this exciting new physical phenomena is induced by the 
breaking of the inversion symmetry,  ${\bf r \rightarrow -r}$, at the
interface (or surface) itself. Also, it is important to notice that by
reducing the spatial dimensionality, to two dimensions (2D) in the 
present case, usually enhances the effects of electron correlations as
it is well-known in fact by studies on cuprates and other layered 
perovskites.
As a result of the broken inversion symmetry, and due to relativistic
considerations, it appears the so-called Rashba effect\cite{rashba},
which describes various momentum-dependent spin-splitting processes,
including spin currents and the spin-Hall
effect.\cite{dyakonovperel,hirsch} The relativistic or Rashba
spin-orbit coupling (RSOC) is usually stated by the
Hamiltonian:\cite{winkler}
\begin{eqnarray}
H_{SOC} = \alpha_R ({\bf k} \times {\bf \sigma}) \cdot \hat{z}
\label{hamsoi}
\end{eqnarray}
where ${\bf k}$ is the electron momentum, ${\bf \sigma}$ the electron
spin, and $\hat{z}$ is the unit vector normal to the surface or
interface. As emphasized in a rather extensive literature, the Rashba
SOC opens new avenues for applications in spintronic 
devices.\cite{wolf,prinz,awschalom} The Rashba
coefficient $\alpha_R$ is in principle proportional to the electric
field appearing due to the broken inversion symmetry, and in an
appropriate device, this electric field can also be tuned by an
external gate voltage.

Various types of interfaces between TMOs where RSOC is present have been 
studied, mostly involving SrTiO$_3$\cite{caviglia,benshalom,ykim,caprara},
particularly LaAlO$_3$/SrTiO$_3$\cite{banerjee,gopinadhan,sahin}, but
also LaMnO$_3$/SrMnO$_3$ interfaces\cite{hwang} have been considered.
The contribution of gate tunable RSOC in addition
to the in-plane RSOC of broken symmetry origin, has been measured in
LaAlO$_3$/SrTiO$_3$ interfaces.\cite{caprara,gopinadhan}
A Rashba SOC has also been reported in devices involving magnetic
layers such as a Co layer with asymmetric Pt and AlO$_x$
interfaces\cite{miron,xwang,pesin}.
Experimental indications of Rashba SOC have also
been reported at surfaces in SrTiO$_3$\cite{nakamura} and 
in KTaO$_3$\cite{pdcking}.

The microscopic description of several transition metal oxides and heavy
fermions
is achieved through generalizations of the ferromagnetic Kondo lattice
model\cite{zener,degennes,hallberg,vandenbrink} (FKLM) also called the 
double exchange model particularly when the Hund exchange coupling is
much larger than the hopping integral. Renewed features and
materials to which this model can be applied are discussed in
Ref.~\onlinecite{qimiaosi}.
For a large Hund coupling, it is known that a metallic ferromagnetic
(FM) phase up to a filling $\nu\sim 0.8$, followed by a tendency towards
an antiferromagnetic (AFM) state with semiconducting or insulating
properties.\cite{dagotto,yunokiPRL98,moreo1999,
mangarep,motome,ishizuka,chatto,santos} Another issue that has
been thoroughly discussed in the context of manganites is the presence
of an instability towards phase separation.\cite{mangarep,dagotto-ps}
Most of these previous studies have been accomplished using the highly
simplifying hypothesis that the electrons in the localized orbitals 
behave as classical continuum spins, which is a reasonable assumption at 
least for manganites.\cite{note1} This assumption allows the use of a 
finite-temperature
Monte Carlo technique to sample the classical spins.\cite{yunokiPRL98}

The main goal of the present study is to determine how the Rashba
spin-orbit coupling affects the magnetic and transport properties in
the 2D FKLM, particularly in its simplest form when only one conduction
electron orbital is included. Although a quantitative study of
interfaces in TMOs and hence a comparison with for example transport
experiments\cite{caviglia} would require a multi-orbital
model,\cite{daghofer} we believe that a first effort
to qualitatively understand the effects of the RSOC should start 
from the single-orbital FKLM, following the program which has been
pursued for example in the study of manganites.
In fact, as we will show below, the resulting model Rashba-FKLM 
presents a highly complex interplay between magnetic and transport
properties which deserves a careful study before taking into account
more involved multi-orbital models. Due to the increased complexity implied
by the inclusion of the RSOC, we also concentrate our study to zero
temperature, although actually most of the
previous studies have dealt with essentially zero-temperature properties.
This allows us to study the RSOC in the full range from zero to infinity,
with respect to the band hopping parameter. The issue of phase separation
will also be addressed in the present effort since experimental 
indications for this instability on interfaces with strong
Rashba SOC have also been reported.\cite{luli}

The Kondo lattice model supplemented by a Rashba spin-orbit coupling
for the conduction electrons on the hexagonal lattice has been recently
considered to study topological and Kondo insulating phases.\cite{xyfeng}
We would also like to stress that the relativistic Rashba SOC is not
the on-site or atomic spin-orbit interaction usually discussed in the
context of TbMnO$_3$\cite{hu2007}, Sr$_2$IrO$_4$\cite{watanabe}
or in the perovskite-like Ruddlesden-Popper series of the ruthenates
Sr$_{n+1}$Ru$_{n}$O$_{3n+1}$.\cite{behrmann} This
is also the case of recently studied models for the LaAlO$_3$/SrTiO$_3$
interface.\cite{ykim} In the model considered in the present work,
the RSOC contributes to the kinetic energy and competes with
spin-conserving hopping term.\cite{riera} We would like to emphasize
that the above mentioned systems where the RSOC takes place in
ferromagnetic layers\cite{miron,xwang,pesin} can also be modelled by
the Rashba-FKLM with the localized classical spins playing the role of
FM moments. Finally, materials with strong RSOC also display many other
interesting features such as the above mentioned topological 
insulators\cite{hasan,andotopins,zhang}, and relativistic Dirac
electrons in graphene\cite{castroneto} and other
compounds.\cite{huynh,mnali}

The paper is organized as follows. In Sec.~\ref{smodel}, we introduce
our Rashba-Ferromagnetic Kondo Lattice model. The zero temperature
perturbative MC method is described in Sec.~\ref{smethod}. Results
for magnetic and transport properties, and phase separation, for all 
electron fillings are
presented in Sec.~\ref{sres}, together with a study of the effect of
magnetic fields and spectral functions which are performed only at
quarter filling.

\section{Model}
\label{smodel}

It is well-known that in TMOs with perovskite structure,
the originally five-fold degenerate 3d orbitals are split into 
three-fold degenerate $t_{2g}$ orbitals $xy, yz, zx$ and two-fold 
degenerate $e_{g}$ orbitals $x^2-y^2, 3z^2-r^2$.
The prototypical TMO compounds where the FKLM was applied are
the manganites with occupied $t_{2g}$ orbitals that appear as 
localized spins. However in the surface/interfaces of this material
the RSOC is somewhat weak, although it has been suggested that
the spin-spiral state of orthorhombic manganites is strongly
deformed by their relativistic spin-orbit interaction.\cite{solovyev}
In SrTiO$_3$, only the $t_{2g}$ orbitals are partially occupied and
involved in both hopping and Rashba processes.
However, at the surface/interface, the filling of orbitals could
change due to orbital mixing.\cite{khalsa,bucheli}
More importantly, itinerant electrons in the $t_{2g}$ bands
of Ti interact with local magnetic moments originating from
electrons localized at the interface,\cite{banerjee,ruhman} which
could be described by a Rashba-FKLM as the one here studied.
Besides, across an interface, located in the $xy$ plane, due to
symmetry constraints, the solely surviving hoppings are those
involving $zx/zx$, $yz/yz$, and $3z^2-r^2/3z^2-r^2$
orbitals.\cite{hwang} There is additional complexity at the 
interface such as dynamical transfer of electrons from the bulk and
location of the interface layer.\cite{gopinadhan,shanavas}
A possible effective Rashba coupling has been proposed for 
Sr$_2$RuO$_4$, where a FKLM for its d$^4$ orbitals would be 
appropriate, only within its chiral superconducting state.\cite{sau}
Finally, the model studied in the present work could be applied
to describe the observed tilting of magnetic order in the 
already mentioned systems containing a Co layer with asymmetric
interfaces.\cite{miron,xwang,pesin} In this case, the interaction
between localized spins $J$ (see below) should be strongly ferromagnetic.

Hence, we believe that important insights on the physics of 
perovskite interfaces can be obtained from the Rashba-FKLM for a single
delocalized orbital coupled to classical localized spins defined
by the Hamiltonian:
\begin{eqnarray}
H_{1o} &=& H_{0} + H_{int}   \nonumber  \\
H_{0} &=& - t_0 \sum_{<l,m>,\sigma} (c_{l\sigma}^\dagger c_{m\sigma} +
    H. c.) + \lambda_{SO} \sum_{l}
     [c_{l+x\downarrow}^\dagger c_{l\uparrow}  \nonumber  \\
  &-& c_{l+x\uparrow}^\dagger c_{l\downarrow} + i (
   c_{l+y\downarrow}^\dagger c_{l\uparrow}
  + c_{l+y\uparrow}^\dagger c_{l\downarrow}) + H. c.] \nonumber  \\
H_{int} &=& - J_H \sum_{l} {\bf S}_l \cdot {\bf s}_l +
          U \sum_{l} n_{l\uparrow} n_{l\downarrow}  \nonumber  \\
        &+& J \sum_{<l,m>} {\bf S}_l \cdot {\bf S}_m
\label{ham1orb}
\end{eqnarray}
The first term in the noninteracting part $H_0$, is the usual hopping
term, $H_{0,hop}$, and the second one corresponds to the RSOC,
$H_{0,SO}$, assuming a square lattice in the $xy$ plane ($z$ is the spin
quantization axis).\cite{pareek} The first term in the interacting part
of the Hamiltonian $H_{int}$ is the ferromagnetic Hund term, $H_H$,
between localized ${\bf S}_l$ and conduction electron ${\bf s}_l$ spins.
The second term is 
the Hubbard repulsion between conduction electrons, $H_U$, and the last
one corresponds to the antiferromagnetic exchange Hamiltonian between
localized spins, $H_J$. This last term is due to virtual processes
involving various hoppings and the Coulomb interaction $U$, and it
may be antiferromagnetic or ferromagnetic.
The notation for the coupling $-J_H$ in $H_{int}$ is drawn from the
Kondo lattice model and has been widely used for the one-orbital FKLM
and even in the two-orbital double exchange model.\cite{vandenbrink}
In Kanamori's notation, it should read $-2 J_H$, but this notation 
is mostly used in multi-orbital models for TMO.
As noticed in previous literature,\cite{yunokiPRL98,dagotto-ps} a
large value of the Hund coupling is appropriate for manganites.
We should emphasize however that the model here proposed, as above 
discussed, is not solely proposed for manganites but for various other
compounds and devices. In this sense, the important issue is that the
adopted value of $J_H$ (see below) leads to the presence of a broad
FM region, followed by a phase separated one with dominant AFM
correlations, which are the main phases of the 2D FKLM.

To start the study of the effects of the RSOC on the known properties
of the FKLM in 2D, we set $U=0$ and $J=0$. Actually, it is well-known 
that a large $J_H$ prevents double occupancy in one-orbital models so
in principle $U=0$ is not very restrictive. In addition, it has been
shown that FM correlations at intermediate fillings, which are 
already present for not very large $J_H$, are just enhanced by 
$U$.\cite{hallberg,dagotto} The effect of $U$ at large fillings on phase
separation is more involved. It would be expected that a finite $U$
would prevent the tendency to PS but actual calculations have shown that
this only occurs for $U\gg J_H$,\cite{mangarep} which is a situation
outside of our current interest. Interestingly, it was shown that for a
spinless two-orbital ``Hubbard" model, valid for the FM phases and
large-$J_H$ limit of manganites, the kinetic energy of a much simpler
single-band model can be used to mimic the one of the e$_g$ 
model.\cite{feiner} It is also important to realize that the case
$U=0$ is also the relevant one for three-orbital models for SrTiO$_3$ 
surfaces and interfaces.\cite{khalsa,bucheli,ruhman}

We normalize the hopping and RSOC
parameters as $t^2_0+\lambda^2_{SO}=1$ whose square root will be
henceforth adopted as the unit of energy. The RSOC implies
the movement of electrons and hence it has a kinetic energy associated 
with it. Then, this normalization keeps the {\em total} kinetic energy
approximately constant with $\lambda_{SO}/t_0$ as is shown in 
Subsection~\ref{transpro}. The ratio $\lambda_{SO}/t_0$ can be considered
as the tangent of the angle between the spin-flipping and spin-conserved
hoppings. The whole purpose of this normalizaion is to keep constant the
ratio between $J_H$ and the total kinetic energy, for a fixed $J_H$.
Alternatively $t_0$ could be kept fixed and adopted as the unit of
energy, but in this case a change in $\lambda_{SO}$ would imply an
effective change of $J_H$ because the total kinetic energy would also
change. In this case we would have to deal with the double effect of
varying $\lambda_{SO}/t_0$ {\em and} $J_{H,eff}/t_0$ which would make
the analysis less clear. In any case, we have verified that the 
results do not change qualitatively with both conventions in the range 
$\lambda_{SO} \leq t_0$.

We adopted throughout the value of $J_H = 10$ in this 
unit, which satisfies the above mentioned requirements, as shown 
in Subsection~\ref{sres}.
The solely parameters left are then the ratio
$\lambda_{SO}/t_0$, and the electron filling $\nu \equiv N_e/N$
($N\equiv L \times L$).

\section{Method}
\label{smethod}

In this work, we will employ a Monte Carlo technique that is
based on the assumption that the localized spins are described by
classical continuum spins ${\bf S}_l=(S, \theta_l, \phi_l)$ in spherical
coordinates.\cite{dagotto,yunokiPRL98,moreo1999}

The technique works at follows. Starting from a given set of 
$\theta_l,\phi_l$, a new configuration is proposed by changing at a
given site $j$, $\theta_j \rightarrow \theta'_j=\theta_j+\Delta \theta$,
$\phi_j \rightarrow \phi'_j=\phi_j+\Delta \phi$. The new configuration
is accepted by computing the difference in total energy
$\Delta E= E(\theta',\phi')-E(\theta,\phi)$ with the usual Metropolis
(or Glauber) criterion. For the noninteracting case ($U=0$ in the 
one-orbital Hamiltonian), the single-particle Hamiltonian is 
diagonalized using library subroutines and the ground state 
$|\Psi_0\rangle$ is built by filling the lowest $N_e$ states ($N_e$ is
the total number of electrons, see Appendix~\ref{spformal}).
For the interacting case, a full diagonalization of the many-body 
problem requires the implementation of for example the Lanczos algorithm
which is much costlier than the treatment of the single-particle
problem and severely restricts the size of the clusters that can be
studied. In both cases, interacting and noninteracting, to diagonalize
the Hamiltonian at each site is excessively expensive and we resort
to the so-called perturbative Monte Carlo 
(PMC)\cite{duane,troung,kennett},
in which the full Hamiltonian is only diagonalized after a sweep on the
whole lattice.

In this perturbative version, at each sweep a site is chosen
(sequentially in the present work) and a change is proposed,
$\theta_j \rightarrow \theta'_j$, and $\phi_j \rightarrow \phi'_j$. Now,
the difference in energy is computed as 
$\Delta E=\langle \Psi_0|\Delta H|\Psi_0 \rangle$, where $\Delta H$
involves only local changes in the Hamiltonian, more specifically in the
terms $H_H$ and $H_J$, which are derived in Appendix~\ref{varham}.

For completeness, we would like to emphasize that the present MC
technique is {\em classical}. It should not be confused with a quantum
MC (QMC) technique as for example the finite-temperature path-integral
auxiliary field algorithm applied in Ref.~\onlinecite{capponi}
to the 2D Kondo lattice model. The advantage of this QMC technique is
that it could deal with finite Hubbard repulsion $U$ as well as with
quantum localized spins. Its main disadvantage is that it is
affected by the ``minus sign problem" which renders this technique
virtually inapplicable to models with fermionic degrees of freedom
except when it is performed at half-filling as implemented in
Ref.~\onlinecite{capponi}. In addition to the above mentioned papers,
the classical MC has been widely used for a variety of related 
models, for example to study the double-exchange model on the
pyrochlore lattice.\cite{motome,ishizuka}

Since in this work we limit ourselves to study zero temperature 
properties, the Monte Carlo simulation is reduced to a simple
optimization procedure, the so-called simulated annealed optimization,
in which a Boltzmann weight is used with a parameter that plays the
role of ``temperature". Hence, using the resulting zero-temperature PMC
technique, clusters as large as $16 \times 16$ can be studied with
conventional desktop computers, although
most of the results presented below correspond to the $8 \times 8$
clusters and in some cases we show results for the $12 \times 12$
cluster, just to show that finite size effects are relatively small.
The ``temperature" in the simulated annealing process is lowered from
0.01 to 0.0008 in 8-10 steps, involving a total of one million MC
sweeps with 150,000 MC sweeps in the final measurement stage.
At least two independent runs were performed for each set of
parameters. The energy for the lowest value of the ``temperature"
considered differs from the one of the previous value within a relative
tolerance of $5~10^{-5}$ and, in the worst case, it is enough to achieve
a relative error of 10$^{-4}$ with respect to the exact energy of the FM
state starting from a random configuration, close to quarter filling. 
We have checked the results obtained by the full and the perturbative MC 
for various electron densities and values of $\lambda_{SO}/t_0$, and for
various types of boundary conditions.

We have also taken into account different types of boundary conditions
(BC). Most of the results reported in the following were obtained for open
BC, which avoids problems associated with open and close level shells.
To compute transport properties we adopted mixed BC, open in one
direction and periodic in the other, which would correspond to a closed
``strip" or ring geometry. Results obtained for fully periodic BC, 
although noisier, agrees reasonably well with the ones obtained with
open or mixed BC.

Typical error bars due to PMC statistics are of the size of the symbols 
employed. Results for the Drude peak strongly depend on the boundary
conditions employed and hence the total error for these results are
much larger than the PMC errors, as indicated in 
Subsection~\ref{transpro}.

\section{Results}

\subsection{Magnetic properties}
\label{sres}

\begin{figure}
\includegraphics[width=0.9\columnwidth,angle=0]{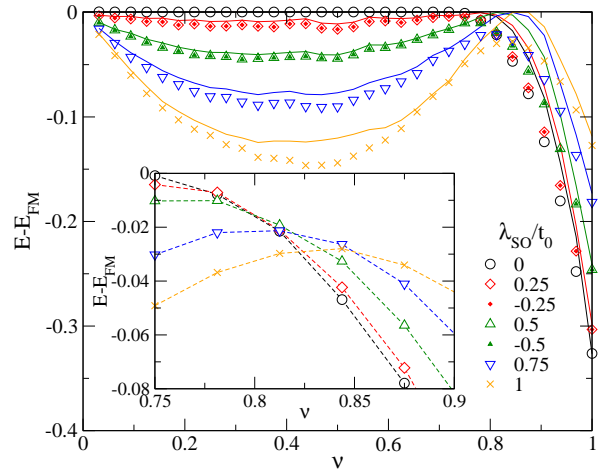}
\caption{(Color online) Energy relative to the energy of the FM state,
per site, 
vs. electron density for various values of $\lambda_{SO}/t_0$. Symbols
are PMC results, lines correspond to trial spiral order. The inset is a
zoom of the $[0.75,0.9]$ density region for the PMC results only.
Results for the $8\times 8$ cluster with open BC.}
\label{fig1}
\end{figure}

In the pure FKLM, various types of magnetic order have been detected
in the $J_H$-density phase diagram in 2D, and this variety is even 
richer in the presence of an AFM interaction between localized
spins $J$.\cite{yunokiPRL98,moreo1999} As we show in this
Subsection, this is also
the case when $\lambda_{SO}$ is turned on, even at a fixed $J_H=10$
and at zero $J$.

Let us start by examining the total energy as a function of electron
density and $\lambda_{SO}$. In Fig.~\ref{fig1}, we show the difference
between the energy and the energy of the FM state, which is exactly
computed by setting $\theta_l=0$ in (\ref{ham1orb}), as a function of
$\nu$ and for various values of $\lambda_{SO}$, obtained by PMC for the
$8 \times 8$ cluster with open BC. For $\lambda_{SO}=0$, the behaviour of
the energy is consistent with the known result of a FM order up to
$\nu \lesssim 0.8$. Now, an increasing RSOC makes the energy to
increasingly depart from the FM level indicating a departure from
the FM order which can be understood as a lowering of the spin-flipping
term $H_{0,SO}$ by an AFM localized spin background.
This departure in energy is maximal close
to quarter-filling ($\nu=0.5$). For larger fillings, for $\lambda_{SO}=0$
the energy strongly departs from the FM one indicating the proximity to
an AFM order that appears due to an effective AFM exchange interaction
caused by virtual processes involving $t_0$. On the other hand, for a
finite $\lambdạ̣_{SO}$, the energy starts to get back closer
to the one of the FM state in spite of the fact that the peak of the
magnetic structure factor (discussed below) continues moving away
from $(0,0)$. Notably, close to half-filling, the effect of the RSOC is
to {\em reduce} the tendency to the AFM state. Evidently, the effective
exchange due to virtual processes involving $\lambda_{SO}$ would no 
longer be AFM.

\begin{figure}
\includegraphics[width=0.9\columnwidth,angle=0]{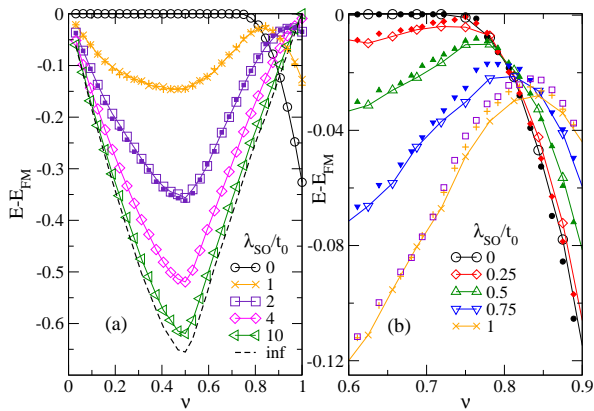}
\caption{(Color online) Energy relative to the energy of the FM state
vs. electron density, per site for various values of $\lambda_{SO}/t_0$.
(a) $\lambda_{SO}/t_0 \geq 1$ ($\lambda_{SO}=0$ added for completeness),
$8\times 8$ cluster.  Lines and open symbols were obtained for open BC; 
stars ($\lambda_{SO}/t_0=1$) and filled squares ($\lambda_{SO}/t_0=2$)
for mixed BC. (b) Comparison of results obtained for the $8\times 8$ 
(lines and open symbols) and $12\times 12$ (filled symbols) clusters with
open BC in the range $0.6 < \nu < 0.9$. For $\lambda_{SO}/t_0=1$,
results for the $12\times 12$ cluster with mixed BC (circles) are also
included.}
\label{fig2}
\end{figure}

It is interesting also to compare the energies obtained by PMC, with the
ones obtained for a fixed localized magnetic order that can be 
plugged in (\ref{ham1orb}) and readily computed. In particular, in most
of the phase diagram, energies quite close to the PMC ones can be 
obtained for a generalized spiral order, defined for the classical
spin at site $(x,y)$ as 
$\theta_{xy}= k_x x + k_y y$, $\phi_{xy}=\phi_0$. In particular
we have examined the values $k_\alpha=2n_\alpha\pi/L$,
$n_\alpha=0,...,L$ ($\alpha=x,y$), and $\phi_0=m \pi/4$, $m=0,...,7$.
Diagonal spiral states, $k_x=k_y$, were mostly previously 
considered,\cite{inoue}
but in the presence of RSOC we found in a large portion of the phase
diagram more stable off-diagonal spiral orders as discussed below.
Of course the FM (AFM) state
correspond to the diagonal spiral state with $k_x=k_y=0$
($k_x=k_y=\pi$). For $\lambda_{SO}=0$ all
values of $\phi_0$ are equivalent, and the energy for a given 
$(k_x,k_y)$ is the same for all the symmetry-equivalent 
${\bf k}$ points. On the other hand, for $\lambda_{SO}>0$, the optimal
energy of the spiral with $(k_x,k_y,\phi_0)$ is degenerate with the one
with $(-k_x,-k_y,\phi_0+\pi)$ but they are different than the optimal
energies for the spiral orders with $(k_x,-k_y,\phi_0+\pi/2)$ and 
$(-k_x,k_y,\phi_0+3\pi/2))$. This symmetry breaking corresponds to the
one observed for the magnetic structure factor as discussed below.
The energy of the spiral states have been added to Fig.\ref{fig1} for
comparison. These energies follow the general trend of the ones 
obtained by PMC.
However, the tendency to FM order observed for $\nu \gtrsim 0.5$ and
$\lambda_{SO} >0$ is actually more pronounced for the spiral state,
and in fact the FM state is actually recovered for 
densities in the
interval $[0.8,0.9]$ in the range of RSOC values considered.

We have also examined the canted state,\cite{degennes} defined by 
$\theta_l=0$ and $\theta_0$ for the two sublattices ($\phi_l=\phi_0$),
but we found that its energies are higher than the ones of the generalized
spiral state for all the parameter space considered, except when both
coincide in the FM or AFM states. This result, for $\lambda_{SO}=0$, is
consistent with previous computational studies for FKLM.\cite{dagotto-ps}

As it can be seen in Fig.~\ref{fig2}(a) the tendency of suppressing the
AFM phase, and approaching to the FM state as the density approaches
half-filling, continues for larger values of $\lambda_{SO}/t_0$, 
including the limiting value $\lambda_{SO}/t_0=\infty$, at which the
FM is finally reached at $\nu=1$. Of course this limit cannot be realized
in real materials but it is of general mathematical interest. In this
plot, results for mixed BC
(or strips) on the same $8\times 8$ cluster are added for comparison
for $\lambda_{SO}/t_0=1$ and 2. The irrelevance of the sign of
$\lambda_{SO}$ has been checked by a set of independent PMC runs.
In Fig.~\ref{fig2}(b) results for
the $8\times 8$ and $12\times 12$ clusters with open BC, and also
for the $12\times 12$ cluster with mixed BC, provide additional
evidence that finite-size effects are negligible for these clusters
and boundary conditions.

To understand the magnetic behavior suggested by the study of the
ground state energies in Figs.~\ref{fig1}-\ref{fig2}, the spin-spin
correlations between localized spins, 
$C({\bf r})=\langle {\bf S}_{\bf r} \cdot {\bf S}_{\bf 0}\rangle$
(${\bf 0}$ is the reference site), and their Fourier
transform leading to the static magnetic structure
function $\chi({\bf k})$, have been computed.

\begin{figure}
\includegraphics[width=\columnwidth,angle=0]{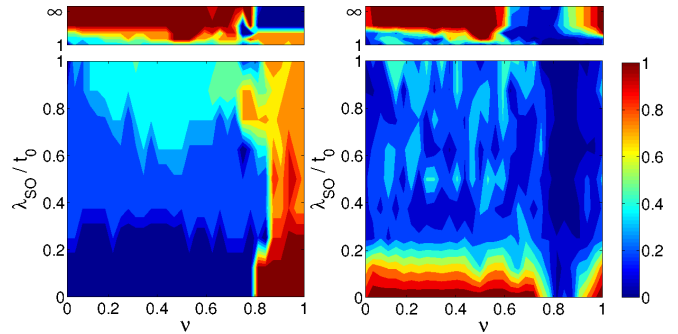}
\caption{(Color online) (a) Magnetic phase diagram in the electron 
density-$\lambda_{SO}/t_0$ plane determined from the modulus of the
peak of the magnetic structure factor, where dark blue
corresponds to FM order ($|{\bf k}_{peak}|=0$) and dark red to
AFM order ($|{\bf k}_{peak}|=1$). (b) Intensity of
$\chi({\bf k}_{peak})$ in density-$\lambda_{SO}/t_0$ plane. Results for
the $8\times 8$ cluster with open BC.}
\label{fig3}
\end{figure}

The RSO coupling leads to a very rich magnetic landscape. To describe
the variety of magnetic orders present let us start by examining the
results for $\chi({\bf k})$, in the density-$\lambda_{SO}/t_0$ plane,
depicted in Fig.~\ref{fig3}. In Fig.~\ref{fig3}(a) we show the modulus of
the peak of $\chi({\bf k})$, $|{\bf k}_{peak}|$, which essentially
describes the proximity to the FM state ($|{\bf k}_{peak}|=0$), or to the
AFM state ($|{\bf k}_{peak}|=1$), its maximum value (in units of
$\sqrt{2}\pi$) for the $8\times 8$ cluster with open BC. For
$\lambda_{SO}=0$ there is a neat crossover from ${\bf k}_{peak}=(0,0)$
to ${\bf k}_{peak}=(\pi,\pi)$ states at $\nu\approx 0.8$, consistently
with previous studies. However, if for $\nu \lesssim 0.8$, $C({\bf r})$
shows an algebraic decay towards a finite value at the maximum distance
indicating long range FM order, for $\nu \gtrsim 0.8$, $C({\bf r})$
indicates short range AFM order except at $\nu =1$ where a long-range
AFM order is achieved. In the $\nu \gtrsim 0.8$ region, the behavior
has been explained as a phase separated AFM-FM state\cite{dagotto-ps}
which we discuss below. In the following the term order will refer to
at least short range magnetic order.

As $\lambda_{SO}/t_0$ is increased, the peak of the magnetic structure
factor departs from both FM and AFM states but remain close to them in
the low and high electron density regions respectively, up to
$\lambda_{SO}/t_0 \sim 2$. It is interesting to note that for very large
values of the RSOC, $\lambda_{SO}/t_0 \gtrsim 4$ the situation is
reversed, that is, the low (high) density region holds now a AFM (FM)
state. One should notice however that, as shown in Fig.~\ref{fig3}(b), the
amplitude of $\chi({\bf k}_{peak})$ is reduced as the RSOC is increased
from zero up to $\lambda_{SO}/t_0=1$, and then it starts to increase again
until recovering its maximum value for $\lambda_{SO}/t_0> 4$.

At quarter filling, where the physics is dominated by the kinetic 
energy, as a difference to half-filling where it is dominated by the
effective AFM exchange interaction, the results of Fig.~\ref{fig3}(a)
suggest an extension of the effective double-exchange model\cite{zener},
\begin{eqnarray}
H_{0,eff} = \sum_{l,m}  (-\tilde{t}_{lm}+\tilde{\lambda}_{l,m}) 
      a_l^\dagger a_m
\label{doubexch}
\end{eqnarray}
where
\begin{eqnarray}
\tilde{t}_{lm}= t_0 \cos{\frac{\theta_{lm}}{2}} \nonumber  \\
\tilde{\lambda}_{l,m}= \lambda_{SO} \sin{\frac{\theta_{lm}}{2}}
\end{eqnarray}
Here $\theta_{lm}$ is the angle between two localized spins at sites
$l$ and $m$. Following Ref.~\onlinecite{izyumov} it is simple to derive this
result at least for the cases when $\theta_l-\theta_m=0$, $\pi$ or $\pi/2$
independently of $\phi_l, \phi_m$.

\begin{figure}
\includegraphics[width=0.9\columnwidth,angle=0]{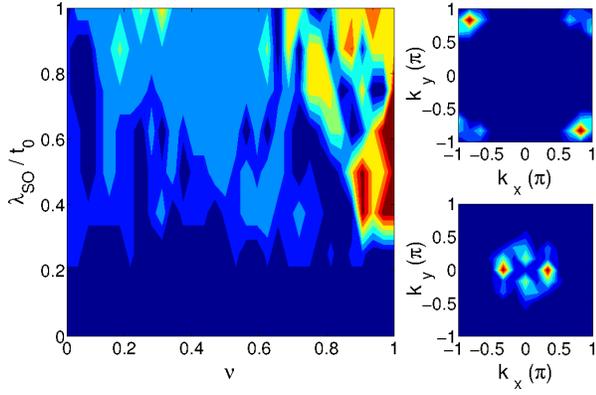}
\caption{(Color online) (a) Magnetic phase diagram in the electron 
density-$\lambda_{SO}/t_0$ plane determined from the distance of the
peak of $\chi({\bf k})$ to the diagonal line in momentum space (see
text), with dark blue (red) corresponding to diagonal (off-diagonal)
${\bf k}$, obtained for the $8\times 8$ cluster with open BC.
Right panels: $\chi({\bf k})$ 
for $\nu=0.97$, (top panel) and $\nu=0.69$ (bottom panel), $12\times 12$
cluster , $\lambda_{SO}/t_0=0.5$, with open BC. The color bar code is the
same as in Fig.~\ref{fig3}}.
\label{fig4}
\end{figure}

It should also be noticed that most of the phase diagram is dominated
by magnetic states characterized by an off-diagonal peak of 
$\chi({\bf k})$, that is, $k_{peak,x}\neq k_{peak,y}$. This behavior is
shown in Fig.~\ref{fig4}, where the distance of ${\bf k}_{peak}$
to the diagonal, $d=|k_{peak,x}-k_{peak,y}|/\sqrt{2}$ is plotted in the
density-$\lambda_{SO}/t_0$ plane. It is apparent that this distance
is maximal for intermediate values of $\lambda_{SO}/t_0$ and close 
to half-filling. It is interesting also to note that for these cases,
the rotational invariance is broken, that is, the PMC simulations are
able to select states where the magnetic structure factor is
maximal only at two points $(k_x,k_y)$, and $(-k_x,-k_y)$ as shown
for two examples, $\nu=0.97$ (top right panel) and $\nu=0.69$ (bottom
right panel), both for $\lambda_{SO}/t_0=0.5$, obtained on the
$12\times 12$ cluster with open BC. This behavior of the magnetic
structure factor indicates the presence of a striped magnetic order as
it was also previously detected for the FKLM in the presence of a
finite exchange $J$ between localized spins or for smaller
$J_H$.\cite{mangarep} Although the energies are slightly higher, there
is in general a good qualitative agreement between the momentum of the
optimal spiral state and the momentum of the peak of $\chi({\bf k})$
obtained by PMC at least for densities $\nu \lesssim 0.75$.

\begin{figure}
\includegraphics[width=\columnwidth,angle=0]{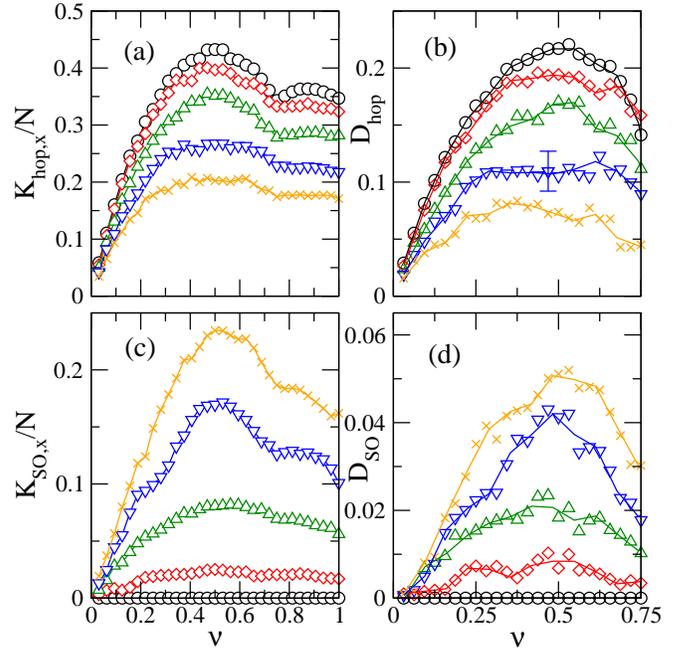}
\caption{(Color online) Kinetic energy per site along $x$ direction
for (a) the spin conserving hopping, and (c) the spin flipping or RSOC
terms of the Hamiltonian as a function of electron density, for various
values of $\lambda_{SO}/t_0$. Drude weight for (b) the spin conserving
hopping and (d) the spin flipping or RSOC as a function of electron
density, for various values of $\lambda_{SO}/t_0$. Symbols and colors
are the same as in Fig.~\ref{fig1}.  Results for the $8\times 8$ cluster
with periodic BC and strips with twisted BC.}
\label{fig5}
\end{figure}

To our knowledge, this is the first report of the influence of the RSOC
on the well-known FM-AFM phases in the 2D FKLM.

\subsection{Transport properties}
\label{transpro}

The optical conductivity is defined as the real part of the linear
response to the electric field and can be written as:\cite{fye}
\begin{eqnarray}
\sigma(\omega) &=& D \delta(\omega) + \sigma^{reg}(\omega)
\label{optcond}  \\
&=& D \delta(\omega)+ \frac{\pi}{L} \sum_{n \neq 0}
    \frac{| \langle \Psi_n | j | \Psi_0 \rangle |^2}{E_n-E_0}
    \delta(\omega - (E_n-E_0))
\nonumber
\end{eqnarray}
where the paramagnetic current along the $x$-direction is:
\begin{eqnarray}
j &=& j_{hop} + j_{SO} \nonumber \\
j_{hop} &=& - i e t_0 \sum_{j, \sigma} ( c_{j+x\sigma}^\dagger c_{j \sigma}
   - c_{j \sigma}^\dagger c_{j+x \sigma} ) \nonumber \\
j_{SO} &=& i e \lambda_{SO} \sum_{j} [(c_{j+x\downarrow}^\dagger
   c_{l\uparrow} - c_{l+x\uparrow}^\dagger c_{l\downarrow}) - H. c.]
\label{current}
\end{eqnarray}
where $j_{hop}$ and $j_{SO}$ are the spin-conserving and spin-flipping
contributions respectively (the electron charge $e=1$ in the following).
The Drude weight $D$ is calculated from the f-sum rule as:
\begin{eqnarray}
\frac{D}{2\pi} = -\frac{\langle H_{0,x} \rangle}{2L} - \frac{1}{L}
\sum_{n \neq 0} \frac{\langle \Psi_n | j | \Psi_0 \rangle |^2}{E_n-E_0}
\label{drude}
\end{eqnarray}
where $K_x \equiv -\langle H_{0,x} \rangle$ is the total kinetic 
energy of electrons along the $x$-direction.

In order to track the contribution from spin-conserving and spin-flipping
transport, from (\ref{optcond}), (\ref{current}) and (\ref{drude}), one 
can formally define the corresponding quantities for the $\lambda_{SO}=0$
and $t_0=0$ limits,
\begin{eqnarray}
\sigma_\alpha(\omega) &=&D_\alpha \delta(\omega)+ \frac{\pi}{L} \sum_{n \neq 0}
    \frac{| \langle \Psi_n | j_\alpha | \Psi_0 \rangle |^2}{E_n-E_0}
    \delta(\omega - (E_n-E_0)) \nonumber \\
\frac{D_\alpha}{2\pi} &=& -\frac{\langle H_{0,_\alpha} \rangle}{2L} - 
    \frac{1}{L} \sum_{n \neq 0} \frac{\langle \Psi_n | j_\alpha |
    \Psi_0 \rangle |^2}{E_n-E_0}
\label{condhopSO}
\end{eqnarray}
with $\alpha=hop, SO$, respectively. Of course, for nonzero $t_0$ and
$\lambda_{SO}$,
$\sigma(\omega) \neq \sigma_{hop}(\omega) + \sigma_{SO}(\omega)$,
$D \neq D_{hop} + D_{SO}$, unless there is no excited state
$|\Psi_n \rangle$ such that
$\langle \Psi_n | j_{hop} | \Psi_0 \rangle\neq 0$ and 
$\langle \Psi_n | j_{SO} | \Psi_0 \rangle \neq 0$
simultaneously.

\begin{figure}
\includegraphics[width=\columnwidth,angle=0]{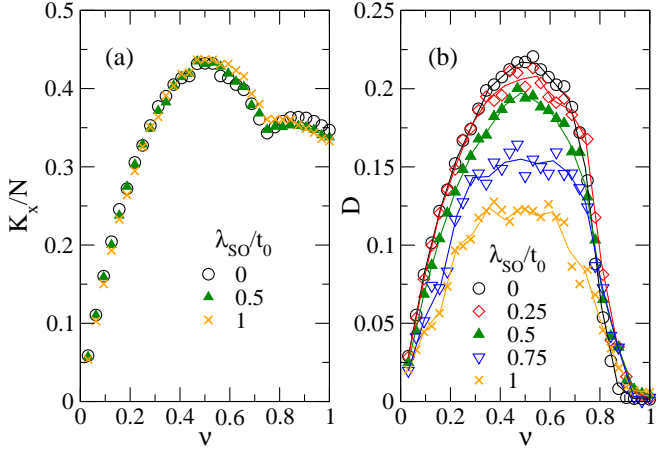}
\caption{(Color online) (a) Total kinetic energy along the $x$ direction,
and (b) total Drude peak for various values of $\lambda_{SO}/t_0$. Lines
are guides to the eye.  Results for the $8\times 8$ cluster averaged over
periodic BC and strips with twisted BC.}
\label{fig6}
\end{figure}

Results for the hopping and SO contributions to the kinetic energy per 
site along the $x$-direction are shown in Figs.~\ref{fig5}(a),(c)
respectively, and results for the corresponding contributions to the
Drude peak, with the above mentioned caveats, in Figs.~\ref{fig5}(d),(b).
These results present oscillations due to various level crossings in spite
of being averaged on the $8 \times 8$ cluster over periodic and strip BC,
in this latter case taking phases enclosing magnetic fluxes equal to 0,
$\pi/2$ and $\pi$ (``twisted" BC). In Fig.~\ref{fig5}(b) we include the
error bars on one point for $\lambda_{SO}/t_0=0.75$ is indicative
of the dispersion of the values obtained for different BC.
Results for the contributions to the
Drude peak are only plotted up to densities $\nu=0.75$, beyond that
the total Drude weight, computed according Eq. (\ref{drude}), starts
to appreciably deviate from $D_{hop} + D_{SO}$. Up to this electron
filling, $D_{hop}$ and $D_{SO}$ roughly follows the behavior of
$K_{hop,x}$ and $K_{SO,x}$ respectively.

\begin{figure}
\includegraphics[width=0.8\columnwidth,angle=0]{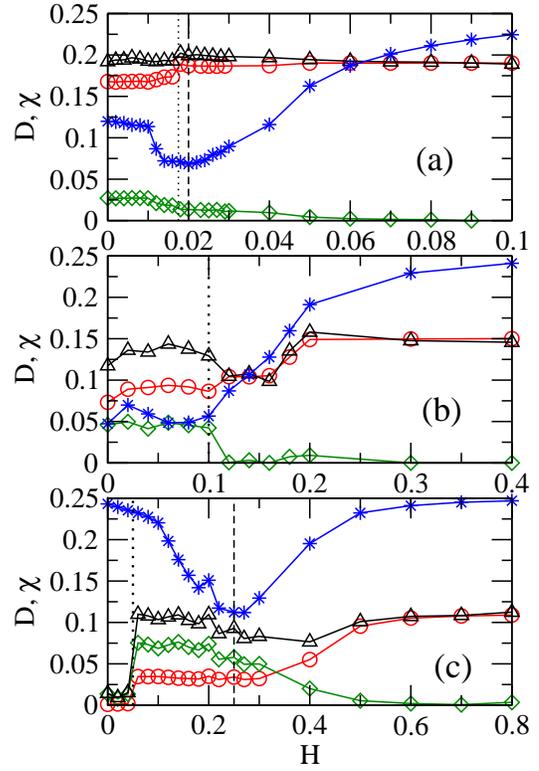}
\caption{(Color online) Drude weight $D$ and maximum value
of the magnetic structure factor $\chi$, for (a) $\lambda_{SO}/t_0=0.5$,
(b) 1.0, and (c) 1.5 at electron density $\nu=0.5$. The hopping and SO
contribution to $D$ are plotted with circles (red) and diamonds 
(green) respectively. Up triangles (black) and stars (blue) correspond
to the total $D$ and $\chi_{max}$ respectively. $\chi$ has been
normalized in this plot in such a way that $\chi_{max}\leq 0.25$.
Vertical dotted and dashed lines correspond to changes
of the momentum of the peak of $\chi$. Results for the $8\times 8$
strip BC.}
\label{fig7}
\end{figure}

The total kinetic energy along the $x$ direction, and the total Drude
peak, calculated using Eq. (\ref{drude}), are shown in 
Fig.~\ref{fig6}. The total kinetic energy (Fig.~\ref{fig6}(a)), for a 
given density, has very small variation in the range
$0 \leq \lambda_{SO}/t_0 \leq 1$. On the other hand, the Drude peak is
suppressed by increasing RSOC in a monotonous way, within the
dispersion of the data previously noticed (Fig.~\ref{fig6}(b)).
For $\lambda_{SO}/t_0 =0$, and for densities $\nu \lesssim 0.75$, the
Drude peak is approximately equal to half the kinetic energy per site,
$K_x/N$, that is, the contribution from
the second term in (\ref{drude}) is very small. It is interesting to
notice a cusp in $K_x/N$ as a function of $\nu$ at $\nu \sim 0.75$
suggesting the presence of a crossover in the transport behavior. For
densities larger than $\nu \sim 0.75$, the Drude peak 
starts to decrease in a rather abrupt fashion, departing from the value
of $K_x/2N$.  By observing the density of states (DOS), this crossover 
suggested by the cusp in  $K_x/N$ vs. $\nu$, could be related to the
Fermi level moving from the bulk of the conduction band towards its
high energy edge as $\nu$ is increased, forming a 
pseudogap.\cite{moreo1999,dagotto-ps} Close to $\nu=1 $ this pseudogap
separates the conduction band and a small peak at slightly higher
energies, which could be considered as an ``impurity" band and hence it
gives support to a semiconductor scenario. These features are well-known
in the pure FKLM\cite{dagotto-ps} and do not change significantly as the
RSOC is turned on.

Another indication of the interplay between magnetic and transport
properties can be observed by computing the optical conductivity in
the presence of a magnetic field $H$ along the $z$-direction, which is 
imposed by adding to the Hamiltonian a Zeeman term,
\begin{eqnarray}
H_{Z}= - H \sum_{l} (S^z_l + s^z_l),
\label{zeeman}
\end{eqnarray}
$\mu_B=\hbar=1$, and the giromagnetic factor has been included in $H$.
Results obtained for the $8\times 8$ cluster with strip BC, at quarter
filling, and for various values of $\lambda_{SO}/t_0$ are shown in
Fig.~\ref{fig7}. In this figure we show the evolution with $H$ of the
hopping and the RSO contributions to the Drude peak, the total Drude
weight, and the maximum value of the static magnetic susceptibility. In
Fig.~\ref{fig7}(a), for $\lambda_{SO}/t_0=0.5$, changes in the Drude
peaks can be observed around $H\sim 0.02$, where two consecutive
crossovers in the peak of $\chi({\bf q})$, first from $(0,\pi/4)$ to
$(0,\pi/2)$ and then to $(0,0)$ also occur. Similarly for
$\lambda_{SO}/t_0=1.0$ (Fig.~\ref{fig7}(b)), changes in the Drude peak
can be observed near $H\sim 0.1$ simultaneously with a change in
$k_{peak}$ from $(0,\pi/4)$ to $(0,0)$. For $\lambda_{SO}/t_0=1.5$
(Fig.~\ref{fig7}(c)) $D_{hop}$, $D_{SO}$ and $D$ change at $H\sim 0.07$
where $k_{peak}$ changes from $(\pi,\pi)$ to $(\pi,0)$, and also at
$H\sim 0.25$ where $k_{peak}$ changes from $(\pi,0)$ to $(0,0)$. It is
clear a general trend at each of those
crossovers of increasing (reducing) the hopping (SO) contribution to the 
Drude peak, although it seems that in most cases the reduction in the
$D_{SO}$ is more important than the increase of $D_{hop}$.
This is understandable since the departure of the FM order due to the
RSOC is precisely opposed by the magnetic field trying to restore the FM
order. We would like to emphasize the fact that the peak of the static
magnetic structure factor does not exhaust the richness of the magnetic
state. In fact, there are in general many other peaks competing with the
one with largest weight. This is the situation for
$\lambda_{SO}/t_0=1.0$, for $0.1 \lesssim H \lesssim 0.2$ where the pair
of peaks $(0,\pi/4)$/$(0,-\pi/4)$ is competing with the dominant one
at $(0,0)$.

\subsection{Phase separation}

\begin{figure}
\minipage{0.55\columnwidth}
\includegraphics[width=\columnwidth,angle=0]{phasep8x8.eps}
\endminipage
\vspace{-0.2cm}
\hspace{0.5cm}\minipage{0.35\columnwidth}
\includegraphics[width=\columnwidth,angle=0]{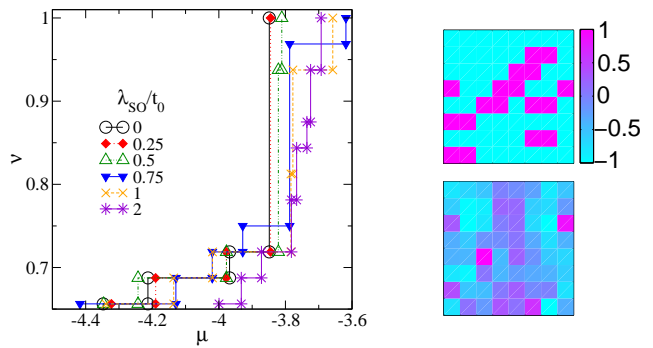}
\endminipage
\caption{(Color online) Left panel: density-chemical potential stability 
diagram for various values of $\lambda_{SO}/t_0$. Right panels: snapshots
of PMC simulations of FM and AFM bonds (see text) for $\lambda_{SO}=0$ 
(top panel) and $\lambda_{SO}/t_0=1$ (bottom panel) at $\nu=0.875$.
Results for the $8\times 8$ cluster with open (left panel) and periodic
(right panels) BC.}
\label{fig8}
\end{figure}

As emphasized in many previous studies on FKLM,\cite{mangarep,dagotto-ps}
the behavior of magnetic and transport properties in the high density
region can be understood by the presence of a phase separated state
between AFM and FM orders. These two orders correspond to different
stable electron fillings, one smaller than $\sim 0.75$ and the other
equal to 1. The electron filling stability is determined by computing
the so-called Maxwell construction, which is performed by adding a
chemical potential term to the Hamiltonian, $-\mu N_e$, where
$N_e$ is the number of electrons, and determining the electron
filling that minimizes the total energy of the resulting Hamiltonian
as a function of the chemical potential $\mu$.

Results for the $8\times 8$ cluster with open BC are shown in
Fig.~\ref{fig8}(left panel) for various values of $\lambda_{SO}/t_0$ in 
the high electron density region. For $\lambda_{SO}=0$, that is, for the
pure FKLM, we recover the well-known phase separated (PS)
state\cite{mangarep,dagotto-ps} which
extends between densities $\nu=0.72$ and 1. As $\lambda_{SO}/t_0$ is
increased such PS state is gradually suppressed, for example for
$\lambda_{SO}/t_0=1$, the largest PS state extends between 
$\nu=0.82$ and $0.94$. This PS region is further reduced by increasing
$\lambda_{SO}/t_0$. Qualitatively similar behavior is obtained for the
$8\times 8$ cluster with periodic and mixed BC, and for the 
$12\times 12$ cluster with open BC.

This suppression of the PS state can be understood by the fact that
the AFM effective exchange, which provides the attractive force leading
to PS, is suppressed by the RSOC, and even at half-filling, the AFM
order disappears for $\lambda_{SO}/t_0 \gtrsim 0.3$ as shown in 
Fig.~\ref{fig3}(a). In fact, the suppression of the PS state is 
actually related to the substitution of the AFM order by some stripe
order with momentum close to $(\pi,\pi)$. This is illustrated in the
right panels of Fig.~\ref{fig8} where snapshots of the values of
${\bf S}_l \cdot {\bf S}_m$ on each horizontal bond during the PMC
simulations, on the $8\times 8$ 
with periodic BC are shown. This quantity is normalized in such a way
that it is equal to 1 ($-1$) for parallel (antiparallel) spins on the
bond. Fig.~\ref{fig8} (top panel) shows a neat PS state between FM and
AFM regions for $\lambda_{SO}=0$, that is the FKLM, consistently with
previous studies. On the other hand, for $\lambda_{SO}/t_0=1$, 
Fig.~\ref{fig8} (bottom panel) shows a very different picture, which
illustrates the suppressed PS and the presence of striped states 
previously discussed.
In any case, by relating the FM domains to conducting ones, and the
close-to-AFM domains to insulating ones, it becomes understandable that
the system behaves for $\nu \gtrsim 0.75$ as a semiconductor
or a poor conductor due to a pseudogap.

\subsection{Spectral functions}

\begin{figure}
\includegraphics[width=0.77\columnwidth,angle=0]{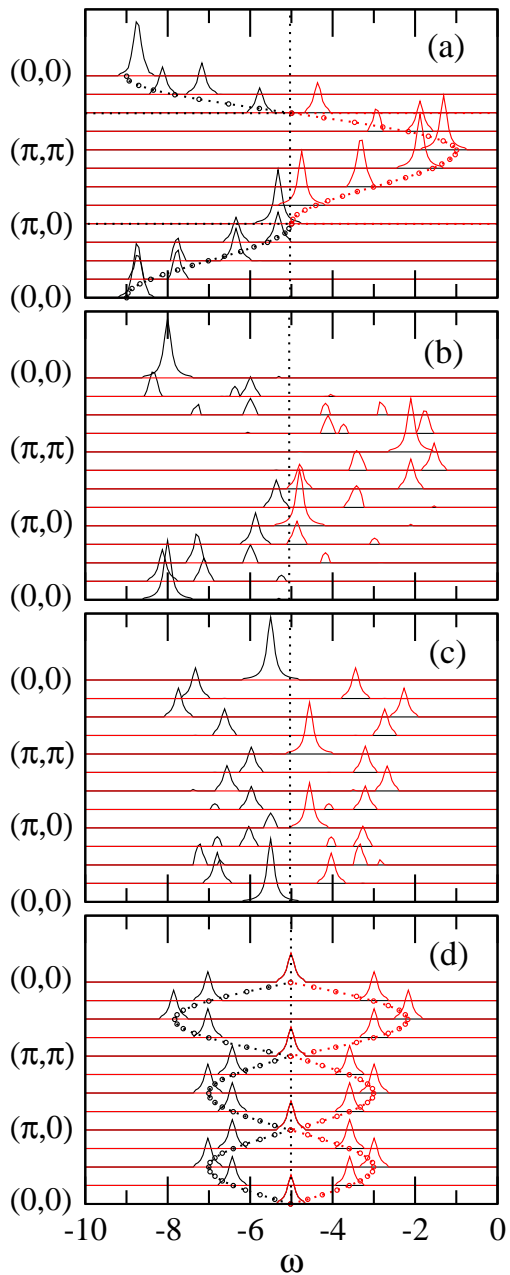}
\caption{(Color online) Spectral functions $A^{(\pm)}({\bf k},\omega)$
for $\lambda_{SO}/t_0=0$ (a), 0.75 (b), 4.0 (c) and $\infty$ (d), at
density $\nu=0.5$ (bonding band). The PES is plotted with full lines
(black) and
the IPES with dashed lines (red). The Fermi level is indicated with
vertical dotted lines. The Lorentzian width employed is $\epsilon=0.1$. 
Results obtained for the $8\times 8$ with periodic BC. Lines with dots
in (a) and (d) correspond to the $32\times 32$ obtained with fixed FM and
AFM order of localized spins respectively.} 
\label{fig9}
\end{figure}

Finally, we studied the spectral functions corresponding to create or
annihilate an electron with momentum ${\bf k}$, 
$A^{(\pm)}({\bf k},\omega)$, defined as:
\begin{eqnarray}
A^{(+)}({\bf k},\omega) = \sum_{n,\sigma}
    | \langle \Psi^{(+1)}_n | c^\dagger_{{\bf k}\sigma} | \Psi_0 \rangle |^2
    \delta(\omega - (E^{(+1)}_n-E_0))  \nonumber  \\
A^{(-)}({\bf k},\omega) = \sum_{n,\sigma}
    | \langle \Psi^{(-1)}_n | c_{{\bf k}\sigma} | \Psi_0 \rangle |^2
    \delta(\omega - (E^{(-1)}_n-E_0))  \nonumber  \\
\label{specfunc}
\end{eqnarray}
$A^{(-)}({\bf k},\omega)$ probes occupied states, and hence it describes 
photoemission spectra (PES), while $A^{(+)}({\bf k},\omega)$ detects
unoccupied levels, and it is conventionally ascribed to ``inverse" PES
(IPES). The $\delta$-peaks are usually considered after being 
broadened by a Lorentzian function with width $\epsilon$.

Fig.~\ref{fig9} shows the lower or bonding band of the spectra (there
is another
identical band shifted at higher energies by $J_H$) at quarter filling
along the line 
$(0,0)\rightarrow (\pi,0)\rightarrow (\pi,\pi)\rightarrow(0,0)$,
obtained for the $8\times 8$ cluster with periodic BC. For the pure
FKLM ($\lambda_{SO}=0$), Fig.~\ref{fig9}(a), the shoulder near
$(\pi,0)$ is responsible for the large conductivity at this filling.
As $\lambda_{SO}/t_0$ increases, the peak at $(0,0)$ ($(\pi,\pi)$)
is shifted to higher (lower) energies thus reducing the number of
states at the Fermi level and consequently the conductivity. This 
reduction of the DOS at the Fermi level could be thought as the 
opening of a pseudogap, but as it is clear from Fig.~\ref{fig9} there
is a restructuring of the Fermi surface as a function of 
$\lambda_{SO}$. In the limit of $\lambda_{SO}/t_0=\infty$, the Fermi
surface only touches the Fermi level at the high-symmetry points
$(0,0)$, $(\pi,0)$, $(0,\pi)$ and $(\pi,\pi)$ of the Brillouin zone
(BZ).

\begin{figure}
\includegraphics[width=0.7\columnwidth,angle=0]{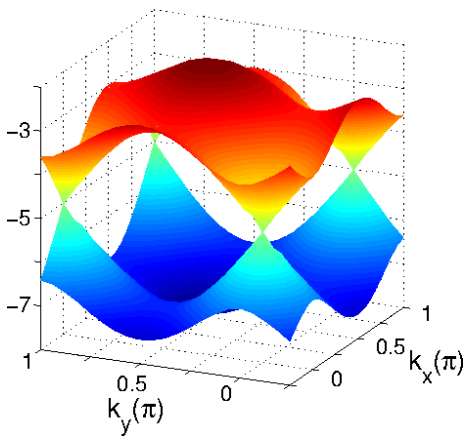}
\caption{(Color online) Energy dispersion for $\lambda_{SO}=1$,
$t_0=0$, at $\nu=0.5$, obtained from $A^{(\pm)}({\bf k},\omega)$.
Results for the $32\times 32$ cluster with AFM fixed
localized spin order, periodic BC.}
\label{fig10}
\end{figure}

For a better visualization of the Fermi surface in the limit of
$\lambda_{SO}/t_0=\infty$, we show in Fig.~\ref{fig10} the energy
dispersion $E^{(\pm)}({\bf k})$ obtained from the first peak above the
Fermi level for $A^{(+)}({\bf k},\omega)$ (upper sheet) and the first
peak below the Fermi level for $A^{(-)}({\bf k},\omega)$ (lower
sheet). Both PES and IPES sheets touch the Fermi level at the $\Gamma$,
$X$, $Y$ and $M$ points, the time-reversal invariant momenta of the
square BZ,\cite{andotopins} consistently with Fig.~\ref{fig9}(d).
The behavior of $E({\bf k})$ at the $\Gamma$, $X$, $Y$ and $M$ points,
resembles the features called Dirac cones, since the linear dispersion
is equivalent to two-dimensional massless Dirac fermions, which are
notably found in graphene\cite{castroneto} but also in other compounds
such as pnictides\cite{huynh} and heavy element compounds.\cite{mnali}
The cones touch the Fermi level, located approximately at 
$\omega=-5.019$, at their vertices.
It is worth to notice that the observed Dirac cones {\em do not} appear
for the noninteracting $H_{0,SO}$ term in Eq.~(\ref{ham1orb}), as it can
be readily checked by a simple tight-binding calculation.

The results in Figs.~\ref{fig9} and \ref{fig10} were obtained for $J=0$,
as all results in this work. Within MC errors, we found the presence of
Dirac cones down to $\lambda_{SO}/t_0=10$ but so far we have not study
this issue systematically. In principle, an AFM $J$ ($J>0$) should
enhance the tendency to an AFM order in the localized spins, and hence
one would expect that the Dirac cones could be realized for lower values
of $\lambda_{SO}/t_0$. Of course, what it really matters is that the
conduction electrons have an AFM order, which implies then that also
the Hund coupling $J_H$ should be large enough. By imposing a fixed AFM
order in the localized spins, for $\lambda_{SO}=1$, $t_0=0$, we actually
found that Dirac cones with vertices at the Fermi level were present for
$J_H$ as low as 3 (in the usual units), at quarter-filling.

\section{Conclusions}

In this work we have analyzed the interplay between the Rashba spin-orbit
coupling on one side, and the hopping and Hund couplings that characterizes
the ferromagnetic Kondo lattice model on the other. Near quarter-filling,
the RSOC moves the 
system away from the ferromagnetic metallic state that is present in the
pure FKLM, leading to a rich variety of magnetic states and to a loss of
conductivity.  Near half-filling, on the other hand, the mechanism that
favours an antiferromagnetic order in the pure FKLM is no longer fully
acting in the presence of the RSOC, and the system presents a tendency
towards striped magnetic orders. As a consequence the presence of phase
separation between antiferromagnetic and ferromagnetic regions in the FKLM
is suppressed by the RSOC. In some studies of the effect of strong Rashba
coupling on transition metal oxides interfaces, it
was reported a tendency towards phase separated state caused by the
RSOC.\cite{caprara} However in the systems considered the RSOC is
proportional to the electron density, a situation that does not
correspond to the model studied in the present work. The system still has
a very low conductivity in the high density region in the presence of a
RSOC, a characteristic of a semiconductor or a pseudogap, since the DOS
is not significantly changed upon switching on the RSOC.

Remarkably, in the limit of RSOC much larger that the hopping integral,
the nature of the magnetic states is reversed, that is at quarter-filling
the system evolves towards an AFM state, and exactly at half-filling the
system becomes a perfect FM.

To our knowledge, this is the first report of the influence of the
Rashba spin-orbit coupling on the well-known ferromagnetic-antiferromagnetic
phases in the 2D ferromagnetic Kondo lattice model, particularly important
when manganites are involved in interfaces.

The general relationship between conductivity and magnetic order 
becomes clear at $\nu=0.5$, particularly when a magnetic field is 
applied through a Zeeman term. Here, it is clear that a more realistic 
model for the orbitals involved in the interface as well as more
details on the interface are needed to make a comparison with
experiments.
 
The profound effects of the RSOC can also be noticed by examining the
spectral functions of creating and annihilating electrons at $\nu=0.5$.
For the pure FKLM, the spectral functions along the main symmetry
lines, resembles that of a Fermi liquid. On the other hand, in the
opposite limit of $\lambda_{SO}\gg t_0$ the Fermi surface is reduced
to Dirac points located at the $\Gamma$, $X$, $Y$ and $M$ points of
the Brillouin zone of the square lattice. The observed Dirac cones are
a nontrivial feature of the RSOC connected to an AFM background by the
Hund coupling.

Future work along this 
direction includes the search of these features for more realistic
sets of parameters better describing the complexity of TMOs
surfaces and interfaces. The rich physics of these interface 
systems, including superconductivity\cite{benshalom} and spin-Hall
effect will also be addressed in future work.

\acknowledgments{We would like to thank Elbio Dagotto for valuable 
comments and suggestions. The authors are partially supported by the
Consejo Nacional de Investigaciones Cient\'ificas y T\'ecnicas (CONICET)
of Argentina.}

\appendix
\section{Hilbert space}
\label{spformal}

The Rashba spin-orbit term, and actually also the Hund's term after
the classical localized spins assumption leading to (\ref{clasHund}),
makes the $z$-projection of the total spin, $S^z_{total}$, no longer
a good quantum number.

In the noninteracting case, in real space, the
single-particle Hamiltonian has to be formulated in the space of spin
up and down electrons, thus becoming a $2N\times 2N$ matrix ($N$ is
the number of cluster sites).

In the interacting case, the Hilbert space has to include all possible
values of $S^z_{total}$, from $-N_e/2$ to $N_e/2$, thus increasing
the difficulty in reaching large cluster sizes. For example, in the
$4\times 4$ cluster, at quarter filling, the dimension of the Hilbert
space is 10,518,300.

\section{Variation of the Hamiltonian}
\label{varham}

It is easy to prove that the Hund coupling between localized and
conduction electron spins can be written as:
\begin{eqnarray}
{\bf S}_l \cdot {\bf s}_{l\alpha} &=&\frac{S}{2} \cos{\theta_l}
(n_{l\alpha\uparrow}- n_{l\alpha\downarrow}) \nonumber \\
&+& S \sin{\theta_l}
(e^{i \phi_l} c_{l\alpha\downarrow}^\dagger c_{l\alpha\uparrow}
+ H. c.)
\label{clasHund}
\end{eqnarray}
In the same way, the AFM coupling between localized spins can be written
as:
\begin{eqnarray}
{\bf S}_l \cdot {\bf S}_m &=& S^2 [\sin{\theta_l} \sin{\theta_m}
\cos(\phi_l - \phi_m) \nonumber \\ 
&+& \cos{\theta_l} \cos{\theta_m}]
\label{classpsp}
\end{eqnarray}

Then, the variation of the $H_H$ term results:
\begin{eqnarray}
\Delta H_H(l) &=& J_H S \sum_{\alpha} \frac{1}{2} 
(\cos{\theta'_l} -\cos{\theta_l})
(n_{l\alpha\uparrow}- n_{l\alpha\downarrow}) + \nonumber \\
&[(&\hskip -0.2cm\sin{\theta'_l} e^{i \phi'_l} -\sin{\theta_l}
e^{i \phi_l}) c_{l\alpha\downarrow}^\dagger c_{l\alpha\uparrow}
+ H. c.)]
\end{eqnarray}

The variation of the $H_J$ term is:
\begin{eqnarray}
\Delta H_J(l) &=& J S^2 \sum_{m(l)} \sin{\theta_m} [\sin{\theta'_l}
\cos(\phi_m - \phi'_l) \nonumber \\
&-& \sin{\theta_l} \cos(\phi_m - \phi_l)] \nonumber \\
&+& \cos{\theta_m} (\cos{\theta'_l}-\cos{\theta_m})
\end{eqnarray}


\begin{thebibliography}{}
\bibitem{note1} At least in one dimension this assumption was checked
      against density matrix-renormalization group calculations using
      $S=3/2$ localized spins in Ref.~\onlinecite{dagotto}.
\end{thebibliography}
\end{document}